# Measuring and Controlling Fishing Capacity for Chinese Inshore Fleets

By

Zheng Yi

Business School, Shanghai Jian Qiao University, Shanghai 201306，China.
Email: yzheng@shou.edu.cn

**Abstract:** The fishing capacity and capacity utilization for Chinese inshore fleets in the latest 13 years were measured by DEA method. Then the relevant models between capacity output, capacity utilization and income were set up, and the function of collecting tax for controlling fishing capacity was quantitative simulated. It pointed out: tax system would be effective for curtailing fishing capacity and improving the efficiency of the whole fishing industry in China, if only the tax rate isn't too low. At last, collecting tax with proper rate was suggested for Chinese inshore fishing fleets.

## 1 Introduction

In marine fishery, the research on fishing capacity is a question for more and more discussion from 1990's to rational collocate capital. The FAO technical working group on the management of fishing capacity defined fishing capacity as the most catch of a vessel or fleet of vessels in definite fish resource and fishing technology during a year or a season. The capacity utilization is actual catch divided by capacity output (FAO, 1998).

Capacity is a potential output. There are basically two ways to measure it: a physical, engineering approach, and the economic approach. In the former, measurement would be achieved by using the technological relationship between the physical attributes of the vessel and output. In the second the measurement would be based on cost minimization.

But how would the measure be able to handle the many-faceted aspects of fishing capacity (vessel, gear, etc)? It is generally believed that accurate and robustness to distribution shifts would be required in decision making processes (Zheng et al., 2022). But the DEA method handles this problem by converting all capital to one measure (through aggregator functions). Secondly, how would one deal with the multi-product issue? Again the DEA method provides a method for dealing with this complication. So the DEA approach can be used to evaluate capacity in either the physical-engineering approach or the economic approach. For these reasons, the experts of FAO recommended the DEA approach emphatically as a method for measuring fishing capacity (FAO, 2000). The technology efficiency (TE) in the DEA approach is as capacity utilization in the research of fishing capacity (Fare etc, 1998).

There is difference for domination character in the management of marine fishery. For studying Chinese management policy of marine fishery, the research range in this paper was limited in the maritime space dominated by Chinese government. It is called by a joint name "Chinese inshore sea" in the following text.

As one of the biggest fishery country, there are the most fishing fleets in China. But the fishing capacity for Chinese inshore fleets is not quantitatively measured as yet. It resulted in the performance of the policy for fishery management can't be effectively evaluated and improved, the number and power of fishing fleets developed too fast since a long period. Because inshore fishing capacity developed blindly, Chinese major economic fish resource declined badly in latest years. It influences the sustainable development of fishing fishery and social stability in the fishing village. So it is essential for Chinese inshore fishery to improve the current management policies and explore new method by exactly measuring its fishing capacity. In the other hand, it has not been researched in the fishery management for simulating the tax policy's function to control fishing capacity.

Base the above case, this paper was firstly going to measure the fishing capacity for Chinese inshore fleets by DEA method. Using the measured results, a serious of simulation models were set up. With these models, the effect to control fishing capacity by collecting tax in Chinese inshore fishing fishery would be quantitative simulated. The results will be as a useful reference for Chinese government to consider setting up some taxing policy with properly tax rate for controlling fishing capacity. At last, collecting tax with proper rate was suggested for Chinese inshore fishing fleets.

## 2　Measuring for Chinese inshore fishing capacity

In this section Chinese inshore fishing capacity in the latest 13 years was measured systemically, so as to provide a basis for the following research on simulating the effect of taxing for Chinese inshore fishing fleets.

After collecting the fishing statistic data of Chinese coastal provinces from 1993 to 2005[1], the inshore fishing capacity and capacity utilization of Chinese coastal provinces were measured by output-oriented DEA method. In this paper, the hypothesis of constant return to scale（CRS）was made and the decision-making units (DMU) in DEA approach are selected as 11 coastal provinces or cities, viz. Tianjing, Hebei, Liaoning, Shanghai, Jiangshu, Zhejiang, Fujian, Sandong, Guangdong, Guangxi and Hainan. The fishing vessels, gross tonnage, total power and labor force in these areas were referred as input indexes and the inshore catch was referred as output index. Making use of Coelli T.J.'s computer program DEAP Version 2.1, the capacity output of all coastal provinces and cities was calculated. After it, they were added up to educe the whole country's inshore capacity output. The calculated results were listed in table 1 and table 2. (The detailed input and output statistic data can be found in appendix.)

Tab.1　The capacity output of 11 inshore provinces or cities
in Chinese sea area from 1993 to 2005 year　($10^4$t）

| Year | 1993 | 1994 | 1995 | 1996 | 1997 | 1998 | 1999 | 2000 | 2001 | 2002 | 2003 | 2004 | 2005 |
|---|---|---|---|---|---|---|---|---|---|---|---|---|---|
| Tianjing | 4.9 | 5.0 | 4.4 | 5.7 | 4.3 | 6.8 | 6.3 | 5.5 | 5.5 | 6.8 | 5.4 | 4.5 | 3.7 |

---

[1] the data in 1993 are rooted in "The statistic collection of Chinese fisheries" (the fisheries bureau, agriculture ministry of China, 1996) , the data from 1994 to 2005 are rooted in the same year's "The statistic yearbook of Chinese fisheries" (the fisheries bureau, agriculture ministry of China, 1995－2006).

| | | | | | | | | | | | | | |
|---|---|---|---|---|---|---|---|---|---|---|---|---|---|
| Hebei | 28.9 | 33.1 | 33.1 | 45.8 | 55.7 | 61.7 | 60.3 | 47.8 | 57.6 | 52.6 | 22.0 | 35.9 | 44.6 |
| Liaoning | 60.3 | 81.3 | 83.5 | 115.6 | 137.2 | 150.3 | 170.2 | 175.7 | 162.6 | 158.5 | 170.5 | 158.3 | 172.9 |
| Shanghai | 7.2 | 8.3 | 8.5 | 9.1 | 8.8 | 7.6 | 7.9 | 6.7 | 6.3 | 6.1 | 11.5 | 4.4 | 4.9 |
| Jiangshu | 56.1 | 69.5 | 81.0 | 125.8 | 147.4 | 151.9 | 146.7 | 133.2 | 110.8 | 95.6 | 41.1 | 96.2 | 89.0 |
| Zhejiang | 165.8 | 232.6 | 288.5 | 322.1 | 321.3 | 335.4 | 337.9 | 319.2 | 311.0 | 302.4 | 302.9 | 293.6 | 291.2 |
| Fujian | 147.2 | 173.3 | 174.6 | 255.6 | 261.1 | 297.4 | 315.2 | 301.2 | 247.7 | 226.9 | 204.8 | 260.8 | 245.4 |
| Sandong | 148.9 | 152.5 | 151.2 | 246.9 | 280.8 | 318.1 | 317.2 | 294.8 | 263.8 | 256.6 | 253.6 | 248.0 | 242.3 |
| Guangdong | 177.9 | 188.2 | 210.8 | 307.9 | 333.4 | 380.8 | 385.6 | 350.8 | 318.4 | 321.9 | 295.1 | 314.5 | 298.6 |
| Guangxi | 34.1 | 39.5 | 49.5 | 70.1 | 78.1 | 88.7 | 88.9 | 88.8 | 89.3 | 86.3 | 85.1 | 79.9 | 84.3 |
| Hainan | 34.1 | 47.1 | 47.1 | 72.2 | 83.2 | 93.6 | 100.3 | 90.1 | 90.7 | 95.3 | 89.4 | 101.6 | 108.0 |
| Whole country | 865.4 | 1030.4 | 1132.2 | 1576.8 | 1711.3 | 1892.5 | 1936.3 | 1813.9 | 1663.7 | 1609.0 | 1481.3 | 1463.3 | 1584.8 |

Tab.2    The capacity utilization of 11 inshore provinces or cities in Chinese sea area from 1993 to 2005 year

| Year | 1993 | 1994 | 1995 | 1996 | 1997 | 1998 | 1999 | 2000 | 2001 | 2002 | 2003 | 2004 | 2005 |
|---|---|---|---|---|---|---|---|---|---|---|---|---|---|
| Tianjing | 0.33 | 0.34 | 0.42 | 0.35 | 0.50 | 0.35 | 0.36 | 0.45 | 0.46 | 0.37 | 0.69 | 0.51 | 0.72 |
| Hebei | 0.44 | 0.40 | 0.47 | 0.46 | 0.46 | 0.49 | 0.54 | 0.68 | 0.55 | 0.60 | 0.46 | 0.85 | 0.69 |
| Liaoning | 1.00 | 0.83 | 1.00 | 0.94 | 0.92 | 0.95 | 0.81 | 0.76 | 0.83 | 0.82 | 0.82 | 0.80 | 0.73 |
| Shanghai | 1.00 | 1.00 | 1.00 | 1.00 | 1.00 | 1.00 | 1.00 | 0.91 | 0.97 | 0.81 | 1.00 | 0.86 | 0.70 |
| Jiangshu | 0.73 | 0.66 | 0.69 | 0.52 | 0.47 | 0.46 | 0.46 | 0.48 | 0.56 | 0.61 | 0.80 | 0.59 | 0.63 |
| Zhejiang | 0.80 | 0.83 | 0.81 | 0.77 | 0.86 | 0.93 | 0.93 | 1.00 | 1.00 | 1.00 | 0.98 | 1.00 | 1.00 |
| Fujian | 0.78 | 0.80 | 0.90 | 0.66 | 0.72 | 0.68 | 0.64 | 0.67 | 0.81 | 0.88 | 1.00 | 0.79 | 0.84 |
| Sandong | 1.00 | 1.00 | 1.00 | 1.00 | 1.00 | 1.00 | 1.00 | 1.00 | 1.00 | 1.00 | 1.00 | 1.00 | 1.00 |
| Guangdong | 0.74 | 0.73 | 0.71 | 0.59 | 0.55 | 0.50 | 0.49 | 0.53 | 0.57 | 0.54 | 0.69 | 0.50 | 0.53 |
| Guangxi | 0.89 | 1.00 | 1.00 | 1.00 | 1.00 | 1.00 | 1.00 | 1.00 | 1.00 | 1.00 | 1.00 | 1.00 | 1.00 |
| Hainan | 0.81 | 0.67 | 0.72 | 0.51 | 0.48 | 0.47 | 0.51 | 0.66 | 0.76 | 0.83 | 1.00 | 0.97 | 1.00 |

Comparing between the fishing capacity and the practical catch for Chinese inshore fleets in latest 13 years was shown in Fig.1. Divided the value of capacity output by practical catch, Chinese every year's inshore capacity utilization was got. It was shown in Fig.2.

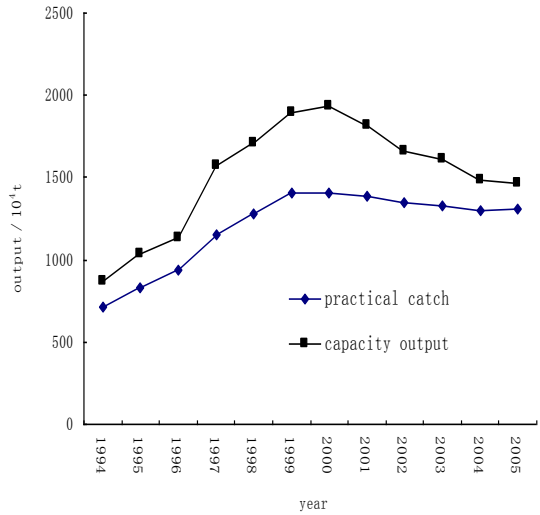 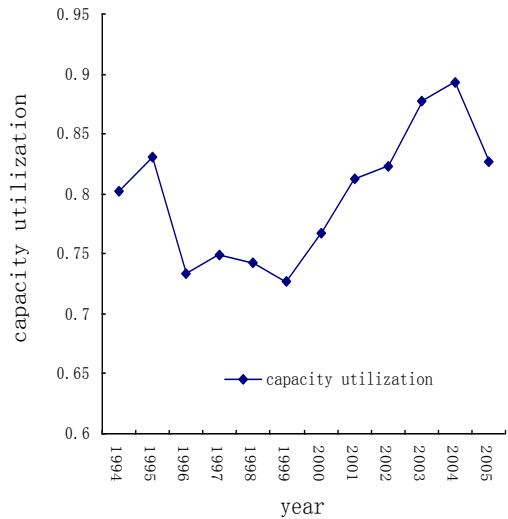

Fig.1 The capacity output and practical catch for chinese inshore fleets

Fig.2 The capacity utilization for Chinese inshore fleets

Analyzing Fig.1, the development of practical catch and capacity output were presented an increasing trend before 1999. From 1999 to 2003 the decreasing appeared. It illustrated that since 1999 Chinese policies for controlling fishing capacity, such as null-increase for marine fishing yield, fishing vessel buying back and fishers' transferring job, produced certain integrated effect. But Fig.1 also shows that the decrease extent was gradually lessening. It pointed out it is more difficult to control fishing capacity.

In Fig.1 we also found: the changing of capacity output was more obvious than practical catch. So as an index of fishery management, capacity output is more sensitive than statistic yield. It can find and settle the problem more early. Analyzing the process of calculating capacity by DEA approach, the major reason of above phenomenon may be that the computation for capacity output not only picked up useful information from statistic yield, but also properly revised some distortion through the integrative analyzing to the input indexes, such as the number of vessels, gross tonnage and total power. So it is very profitable for scientific fishery management to measure fishing capacity by DEA method.

## 3  Set up simulation models

Based the above measuring results, the function of collecting tax would be simulated for controlling inshore fishing capacity. The main idea is: firstly, the general changing rule of Chinese fishing capacity was explored with econometrics models; then the simulation would be done by making use of the models and the statistic data.

For setting up simulation models, the past years' gross income (the fisheries bureau, agriculture ministry of China, 1995－2006) and the related above calculation results ( the last line in Tab.1) were listed in Tab.3.

Tab.3    Fishing's income and output for Chinese inshore fleets from 1994 to 2005 year

| Year | Inshore practical catch (t) A | Inshore fishing capacity (t) B | Increment of inshore fishing capacity (t) C (B-B(-1)) | Income of inshore fishing (yuan) D | Practical income of capacity output to every ton (yuan/t) E (D/B) | income of practical catch to every ton (yuan/t) F (D/A) | Capacity utilization G (A/B) | Increment of Capacity utilize-tion H (G-G(-1)) |
|---|---|---|---|---|---|---|---|---|
| 1993 | 7088632 | 8654471.5 | | | | | 0.819 | |
| 1994 | 8270574 | 10304461 | 1649989.47 | 6826772241 | 662.5 | 825.4 | 0.803 | -0.016 |
| 1995 | 9411578 | 11322073 | 1017611.94 | 8939278809 | 789.5 | 949.8 | 0.831 | 0.029 |
| 1996 | 11563258 | 15767867 | 4445793.94 | 9331585044 | 591.8 | 807.0 | 0.733 | -0.098 |
| 1997 | 12816781 | 17113476 | 1345609.64 | 10038032011 | 586.6 | 783.2 | 0.749 | 0.016 |
| 1998 | 14053674 | 18924631 | 1811154.38 | 11201495847 | 591.9 | 797.1 | 0.743 | -0.006 |
| 1999 | 14077132 | 19362909 | 438277.709 | 11104414113 | 573.5 | 788.8 | 0.727 | -0.016 |
| 2000 | 13909342 | 18138509 | -1224399.4 | 11068689991 | 610.2 | 795.8 | 0.767 | 0.040 |
| 2001 | 13521266 | 16636523 | -1501986 | 11627395978 | 698.9 | 859.9 | 0.813 | 0.046 |
| 2002 | 13238479 | 16089554 | -546968.76 | 11589863179 | 720.3 | 875.5 | 0.823 | 0.010 |
| 2003 | 12992035 | 14813145 | -1276409.4 | 11766145042 | 794.3 | 905.6 | 0.877 | 0.054 |
| 2004 | 13059784 | 14633268 | -179877 | 12517503013 | 849.7 | 958.5 | 0.892 | 0.015 |
| 2005 | 13094900 | 15848054 | 1214785.71 | 13413190885 | 846.4 | 1024.3 | 0.826 | -0.066 |

Making use of the data in Tab.3 and the theory about the model of lagged variable, the relation between capacity output, capacity utilization and income for inshore fishing fleets were set up.

## 3.1 Relation between the fishing capacity and the income

By the help of Almon polynomial method and software of Eviews, the lagged model between the fishing capacity and the income for Chinese inshore fishing fleets was set up as follows:

$$CZL = -4026971.665 + 0.00185D + 0.00006D(-1) - 0.00085D(-2) - 0.00087D(-3) \quad \ldots\ldots(1)$$

Where CZL was represented an increment of inshore fishing capacity, D was an income for inshore fishing fleets, D(-1)、D(-2) and D(-3) was separate denoted the lagged value of the first period, the second period and the third period of the income. Outcome computed by the software of Eviews was shown in tab.4, it means the relativity between the fishing capacity and the income is rather obviously.

Tab.4 The estimation result in lagged model between capacity and income

| Variable | Coefficient | Std. Error | t-Statistic | Prob. |
|---|---|---|---|---|
| C | -4026972. | 1527505. | -2.636307 | 0.0387 |
| PDL01 | 5.53E-05 | 3.83E-05 | 1.442744 | 0.1992 |
| PDL02 | -0.001351 | 0.000135 | -10.02745 | 0.0001 |
| R-squared | 0.954309 | Mean dependent var | | 8909.655 |
| Adjusted R-squared | 0.939079 | S.D. dependent var | | 1248694. |
| S.E. of regression | 308205.6 | Akaike info criterion | | 28.37612 |
| Sum squared resid | 5.70E+11 | Schwarz criterion | | 28.44187 |
| Log likelihood | -124.6926 | F-statistic | | 62.65847 |
| Durbin-Watson stat | 3.140655 | Prob(F-statistic) | | 0.000095 |

Formula (1) shows that: in latest decade years, the changing of Chinese fishing capacity had a close connection with the fishers' income in same and latest three years. It also indicated: if total fishing income in a year was decreased ten thousand yuan, the capacity output in the same year would be decreased about 18.5 ton. So it is a good idea for controlling fishing capacity to control the gross income in inshore fishing fleets. These calculation results accords with the theory of externalities in economics, i.e. the fishing externalities will be decreased by increasing fisher's cost so as to decrease their income. By it the private cost will be consistent with the social cost.

## 3.2  Relation between the capacity utilization and capacity output

Suppose OUT is represented capacity output, its unit is ten thousand ton. OUT(-1) and OUT(-2) are lagged one and two period's capacity output. For lightening the multiple common linearity between the capacity output and its lagged value, the relation between capacity output and the compound variables （OUT—OUT(-1)）and（OUT(-1)—OUT(-2)） was estimated by dualistic regression method as follows:

UZL=0.00835-0.00027(OUT-OUT(-1))+0.00008(OUT(-1)-OUT(-2)) ……..(2)

Where UZL is represented as an increment of capacity utilization. Outcome computed by Eviews was shown in tab.5, it means the relativity is also obviously. In the other hand, the method of composing new variables was validated to be effectively for mitigating the multiple common linearity, because the correlation coefficient of the two compound variables is only 0.486, but the correlation coefficient between capacity output and its lagged value is bigger than 0.9.

Tab.5  Relation between the increment of capacity utilization and capacity output

| Variable | Coefficient | Std. Error | t-Statistic | Prob. |
|---|---|---|---|---|
| C | 0.008351 | 0.007327 | 1.139746 | 0.2919 |
| OUT-OUT(-1) | -0.000273 | 4.63E-09 | -5.903734 | 0.0006 |
| OUT(-1)-OUT(-2) | 0.000081 | 4.65E-09 | 1.746236 | 0.1243 |
| R-squared | 0.839107 | Mean dependent var | | -0.000498 |
| Adjusted R-squared | 0.793137 | S.D. dependent var | | 0.048841 |
| S.E. of regression | 0.022214 | Akaike info criterion | | -4.532864 |
| Sum squared resid | 0.003454 | Schwarz criterion | | -4.442088 |
| Log likelihood | 25.66432 | F-statistic | | 18.25353 |
| Durbin-Watson stat | 1.878494 | Prob(F-statistic) | | 0.001671 |

Formula (2) shows that the changing of capacity utilization was largely decided by the capacity output in the latest three years. It is say that the effect to fishery resource by fishing capacity is much more than the effect by nature.

## 4. Simulation for the function of collecting tax

Because the policy of collecting tax for controlling fishing capacity is not carried out in Chinese fishery now, it is not come true for demonstration. So it is useful for exploring some new policies in fishery management to do the quantitative simulation for the effect of collecting tax.

### 4.1　Simulation method

For simulating the effect of tax to control fishing capacity in the latest decade years, it is supposed that the tax was collected in proportion to practical fishing capacity from 1996 to 2005, and the value of fishing capacity and capacity utilization in 1994, 1995 and 1996, which were calculated by DEA approach in Tab.1 and Tab.2, was served as initial value. According to the income after taxing, the capacity output in every year can be simulated by formula (1). Then capacity utilization can be gained by formula (2). The probably actual catch in the same year can be educed through the capacity utilization multiplying by capacity output.

### 4.2.　Simulation results

By the actual annual statistic data and formulas (1) and (2), the simulation was done. For comparing, the simulated results and actual values were all listed in Tab.6.

Tab.6  Effect for fishing capacity by collecting 100 yuan tax
to a ton capacity outout from 1996 to 2005 year

| Year A | Simulated capacity output after taxing (t) B | Actual capacity output (t) C | Simulated capacity utilization after taxing D | Actual capacity output E | Simulated income of inshore fishing fleets after taxing (yuan) F （G-100*C） | Actual income of inshore fishing fleets after taxing (yuan) G | Simulated income after taxing for a ton's capacity output (yuan/t) H （F/B） | Actual income for a ton's capacity output (yuan/t) I |
|---|---|---|---|---|---|---|---|---|
| 1994 | 10304461 | 10304461 | 0.803 | 0.803 | 6826772241 | 6826772241 | 663 | 662.5 |
| 1995 | 11322073 | 11322073 | 0.831 | 0.831 | 8939278809 | 8939278809 | 790 | 789.5 |
| 1996 | 15767867 | 15767867 | 0.733 | 0.733 | 7754798363 | 9331585044 | 492 | 591.8 |
| 1997 | 14026629 | 17113476 | 0.825 | 0.749 | 8326684365 | 10038032011 | 594 | 586.6 |
| 1998 | 13305826 | 18924631 | 0.839 | 0.743 | 9309032764 | 11201495847 | 700 | 591.9 |
| 1999 | 12922102 | 19362909 | 0.852 | 0.727 | 9168123259 | 11104414113 | 709 | 573.5 |
| 2000 | 11356753 | 18138509 | 0.900 | 0.767 | 9254839080 | 11068689991 | 815 | 610.2 |
| 2001 | 10373492 | 16636523 | 0.923 | 0.813 | 9963743665 | 11627395978 | 961 | 698.9 |
| 2002 | 9509933 | 16089554 | 0.947 | 0.823 | 9980907742 | 11589863179 | 1050 | 720.3 |
| 2003 | 8530580 | 14813145 | 0.975 | 0.877 | 10284830542 | 11766145042 | 1206 | 794.3 |
| 2004 | 8360489 | 14633268 | 0.980 | 0.892 | 11054176213 | 12517503013 | 1322 | 849.7 |
| 2005 | 9392066 | 15848054 | 0.959 | 0.826 | 11828385514 | 13413190885 | 1259 | 846.4 |

Tab.6 indicated: if a hundred yuan tax was collected for a ton's capacity output from 1996 to 2005, the output of fishing capacity in 2005 would be limited to 9,392,065 ton. It is about 59.6％ of the capacity output in the year, when taxing was started (i.e. the year of 1996). But the practical catch in 2005 is a little higher than 1996's. It indicated that the fishing capacity would be curtailed 40% by collecting the tax for ten years, if the other policies for fishery management were not changed in this period. It is say that the effect of the taxing policy should be obviously to control fishing capacity.

In the same time, when the value of D and E column in Tab.6 were compared, the capacity utilization after taxing is obviously bigger than the practical capacity utilization from the next year after taxing, the year of 1997. It is say that the taxing policy may promote the efficiency for whole fishery when it effectively controls fishing capacity, But the actual management methods for controlling fishing capacity apparently had to cut down the efficiency of fishing fleets.

By the same way, the effect for the other tax rate can be simulated. The results were shown in Tab.7.

Tab.7 Analysis about the effect to fishing capacity by different tax rate from 1996 to 2005 year

| Year | Collecting 5 yuan for a ton's capacity output | | Collecting 40 yuan for a ton's capacity output | | Collecting 60 yuan for a ton's capacity output | | Collecting 80 yuan for a ton's capacity output | |
|---|---|---|---|---|---|---|---|---|
| | Capacity output after taxing (t) | Capacity utilization after taxing | Capacity output after taxing (t) | Capacity utilization after taxing | Capacity output after taxing (t) | Capacity utilization after taxing | Capacity output after taxing (t) | Capacity utilization after taxing |
| 1994 | 10304461 | 0.803 | 10304461 | 0.803 | 10304461 | 0.803 | 10304461 | 0.803 |
| 1995 | 11322073 | 0.831 | 11322073 | 0.831 | 11322073 | 0.831 | 11322073 | 0.831 |
| 1996 | 15767867 | 0.733 | 15767867 | 0.733 | 15767867 | 0.733 | 15767867 | 0.733 |
| 1997 | 17118492 | 0.741 | 15979385 | 0.772 | 15328466 | 0.790 | 14677547 | 0.808 |
| 1998 | 18539217 | 0.721 | 16611126 | 0.765 | 15509359 | 0.790 | 14407593 | 0.814 |
| 1999 | 18971259 | 0.729 | 16742622 | 0.775 | 15469115 | 0.801 | 14195609 | 0.826 |
| 2000 | 17750905 | 0.775 | 15395165 | 0.821 | 14049028 | 0.847 | 12702891 | 0.874 |
| 2001 | 16656932 | 0.803 | 14341981 | 0.847 | 13019151 | 0.872 | 11696322 | 0.898 |
| 2002 | 15641443 | 0.830 | 13382466 | 0.873 | 12091621 | 0.898 | 10800777 | 0.922 |
| 2003 | 14505606 | 0.861 | 12304281 | 0.903 | 11046380 | 0.927 | 9788480 | 0.951 |
| 2004 | 14309121 | 0.866 | 12117520 | 0.908 | 10865176 | 0.932 | 9612833 | 0.956 |
| 2005 | 15675451 | 0.835 | 13360520 | 0.881 | 12037702 | 0.907 | 10714884 | 0.933 |
| Cut-tailed extent | 0.6% | | 15.3% | | 23.7% | | 32.0% | |

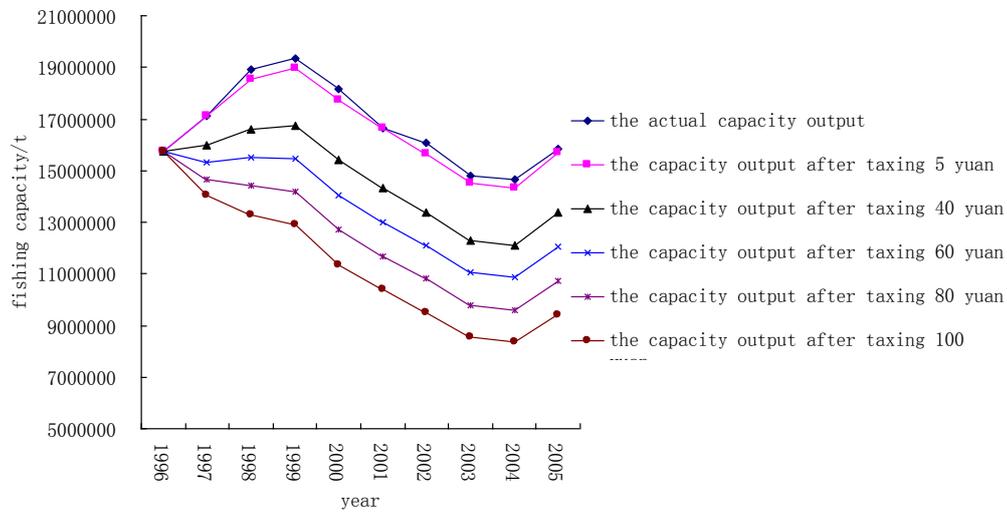

Fig.3 Changing of capacity output by different tax rate

According to Tab.7, the curtailed extent of the simulated fishing capacity in ten years may be 0.6%, 15%, 24% and 32%, when the tax rate was 5, 40, 60 and 80 yuan for a ton's fishing capacity. The changing for every year's fishing capacity with different tax rate was illustrated in Fig.3. It was shown for fishing capacity that the more tax rate the more curtailed extent in ten years. But when the tax rate is low, the difference between fishing capacity and actual catch is small. Such as tax rate was 5 yuan for a ton's fishing capacity, the capacity output after taxing would be almost same as actual catch in Fig.3. The simulation also explained why the actual collected protecting fee for fish resource doesn't work to control fishing capacity in China. Its major reason isn't the fee shouldn't be collected, but the rate is too low.

## 5. Epilogue

In this paper, Chinese fishing capacity for inshore fishing fleets is measured by DEA approach for the first time. Based the calculated results, the relation between fishing capacity, capacity utilization and fishing income was set up. With these models, the effect of taxing in different rate was simulated. According to the simulated results, this paper thought that it is effectively for tax policy to control fishing capacity, but the tax rate shouldn't be too low.

Based the above analyses, taxing for Chinese inshore fishing fleets is thought to be necessary for protecting Chinese fish resource. So this paper suggests to constituting a taxing policy as soon as possible, so as to the fishing tax will be collected with properly rate. It will be helpful for controlling fishing capacity and promoting the efficiency in whole fishery. It also can effectively protect Chinese fish resource.

## References

FAO (1998): Report of the technical working group on the management of fishing capacity. FAO Fisheries Report No. 586, pp. 13.

FAO (2000): Report of the technical working group on the management of fishing capacity. FAO Fisheries Report No. 615, pp. 32-51.

Fare, R. and S. Grosskopf (1998): Primal and Dual DEA Measures of Capacity Utilization. Dept. of Econ. And Dept. of Agri. Res. Economics Discussion Paper, Oregion State University.

The fisheries bureau, agriculture ministry of China, (1995). The statistic yearbook of Chinese fisheries.

The fisheries bureau, agriculture ministry of China, (1996). The statistic yearbook of Chinese fisheries.

The fisheries bureau, agriculture ministry of China, (1997). The statistic yearbook of Chinese fisheries.

The fisheries bureau, agriculture ministry of China, (1998). The statistic yearbook of Chinese fisheries.

The fisheries bureau, agriculture ministry of China, (1999). The statistic yearbook of Chinese fisheries.

The fisheries bureau, agriculture ministry of China, (2000). The statistic yearbook of Chinese fisheries.

The fisheries bureau, agriculture ministry of China, (2001). The statistic yearbook of Chinese fisheries.

The fisheries bureau, agriculture ministry of China, (2002). The statistic yearbook of Chinese fisheries.

The fisheries bureau, agriculture ministry of China, (2003). The statistic yearbook of Chinese fisheries.

The fisheries bureau, agriculture ministry of China, (2004). The statistic yearbook of Chinese fisheries.

The fisheries bureau, agriculture ministry of China, (2005). The statistic yearbook of Chinese fisheries.

The fisheries bureau, agriculture ministry of China, (2006). The statistic yearbook of Chinese fisheries.

Zheng, J., Makar, M. (2022).Causally motivated multi-shortcut identification & removal. Advances in Neural Information Processing Systems 35,12800-12812.

# APPENDIX

Tab.1  Measuring inshore fishing capacity in 1993

| | Inshore fishing caich(t) | Number of engine vessels | | | Labour force | Calculated by DEA | |
|---|---|---|---|---|---|---|---|
| | | Number of vessels | Gross tonnage(t) | Power (kW) | | Capacity utilization | Capacity output(t) |
| Tianjing | 16442 | 936 | 37920 | 56258 | 35882 | 0.334 | 49156.24 |
| Hebei | 126704.5 | 8102 | 116216 | 227749 | 121597 | 0.439 | 288547.5 |
| Liaoning | 603426 | 27259.5 | 252848 | 545174 | 219208 | 1 | 603426 |
| Shanghai | 72446 | 675.5 | 68412.5 | 122821 | 36479 | 1 | 72446 |
| Jiangshu | 410959 | 14176 | 251345.5 | 475713 | 942812 | 0.732 | 561037.5 |
| Zhejiang | 1331700 | 35874 | 1091936 | 2112422 | 745370 | 0.803 | 1657816 |
| Fujian | 1148504 | 50320.5 | 512254 | 1171291 | 776314 | 0.78 | 1472156 |
| Sandong | 1488816 | 42403 | 518051 | 907169 | 624779 | 1 | 1488816 |
| Guangdong | 1310316 | 49784 | 645937.5 | 1584601 | 1173226 | 0.737 | 1778516 |
| Guangxi | 302478 | 9361 | 129872 | 292579 | 391711 | 0.886 | 341234.4 |
| Hainan | 276842 | 12732 | 158243 | 354238 | 133066 | 0.811 | 341320 |
| Whole country | 7088632 | | | | | | 8654471 |

Tab.2　Measuring inshore fishing capacity in 1994

| | Inshore fishing catch(t) | Number of engine vessels | | | Labour force | Calculated by DEA | |
|---|---|---|---|---|---|---|---|
| | | Number of vessels | Gross tonnage(t) | Power (kW) | | Capacity utilization | Capacity output(t) |
| Tianjing | 17143 | 919 | 30092 | 49365 | 2594 | 0.343 | 50008.23 |
| Hebei | 133294 | 8695 | 120200 | 232864 | 30281 | 0.403 | 330685.5 |
| Liaoning | 674002 | 27344 | 259280 | 553635 | 86553 | 0.829 | 812820.8 |
| Shanghai | 83177 | 638 | 54140 | 98772 | 2301 | 1 | 83177 |
| Jiangshu | 455298 | 17337 | 252848 | 482534 | 74993 | 0.655 | 695182 |
| Zhejiang | 1921344 | 37678 | 1184466 | 2239004 | 168044 | 0.826 | 2326473 |
| Fujian | 1382525 | 50467 | 531052 | 1251370 | 212695 | 0.798 | 1733128 |
| Sandong | 1524501 | 43272 | 467126 | 859009 | 175909 | 1 | 1524501 |
| Guangdong | 1370083 | 49532 | 642892 | 1574437 | 192701 | 0.728 | 1881881 |
| Guangxi | 395457 | 9971 | 144336 | 318207 | 29232 | 1 | 395457 |
| Hainan | 313750 | 12669 | 156077 | 364386 | 65665 | 0.666 | 471147.2 |
| Whole country | 8270574 | | | | | | 10304461 |

Tab.3　Measuring inshore fishing capacity in 1995

| | Inshore fishing caich(t) | Number of engine vessels | | | Labour force | Calculated by DEA | |
|---|---|---|---|---|---|---|---|
| | | Number of vessels | Gross tonnage(t) | Power (kW) | | Capacity utilization | Capacity output(t) |
| Tianjing | 18367 | 913 | 22808 | 37545 | 2279 | 0.417 | 44042.69 |
| Hebei | 156718 | 8857 | 117765 | 234208 | 26547 | 0.473 | 331419.1 |
| Liaoning | 834609 | 30565 | 274314 | 592448 | 90957 | 1 | 834609 |
| Shanghai | 84861 | 625 | 54976 | 102583 | 2644 | 1 | 84861 |
| Jiangshu | 558874 | 20299 | 285052 | 567823 | 85992 | 0.69 | 809946.3 |
| Zhejiang | 2347477 | 40216 | 1556439 | 2825791 | 184851 | 0.814 | 2884782 |
| Fujian | 1568951 | 52402 | 567050 | 1454298 | 224604 | 0.898 | 1746236 |
| Sandong | 1511790 | 45589 | 518457 | 999876 | 177915 | 1 | 1511790 |
| Guangdong | 1494157 | 50611 | 684578 | 1761638 | 195146 | 0.709 | 2108165 |
| Guangxi | 495484 | 10418 | 160897 | 362944 | 29073 | 1 | 495484 |
| Hainan | 340290 | 12529 | 152861 | 375806 | 64619 | 0.723 | 470737.1 |
| Whole country | 9411578 | | | | | | 11322073 |

Tab.4   Measuring inshore fishing capacity in 1996

| | Inshore fishing caich(t) | Number of engine vessels | | | Labour force | Calculated by DEA | |
|---|---|---|---|---|---|---|---|
| | | Number of vessels | Gross tonnage(t) | Power (kW) | | Capacity utilization | Capacity output(t) |
| Tianjing | 20144 | 953 | 19578 | 35105 | 2599 | 0.354 | 56909.58 |
| Hebei | 208328 | 8390 | 116239 | 255808 | 25805 | 0.455 | 458028.1 |
| Liaoning | 1086916 | 30413 | 261661 | 595624 | 122731 | 0.94 | 1156161 |
| Shanghai | 91473 | 820 | 64252 | 119944 | 3169 | 1 | 91473 |
| Jiangshu | 653461 | 22072 | 318637 | 642351 | 101263 | 0.519 | 1258024 |
| Zhejiang | 2463247 | 40131 | 1608035 | 3044953 | 196912 | 0.765 | 3220533 |
| Fujian | 1683155 | 54167 | 578562 | 1509021 | 208756 | 0.658 | 2556402 |
| Sandong | 2468575 | 45824 | 558685 | 1123238 | 186059 | 1 | 2468575 |
| Guangdong | 1817253 | 52348 | 758990 | 2037125 | 205865 | 0.59 | 3078992 |
| Guangxi | 700627 | 10934 | 185352 | 431927 | 32001 | 1 | 700627 |
| Hainan | 370079 | 13129 | 167004 | 398111 | 65222 | 0.512 | 722142 |
| Whole country | 11563258 | | | | | | 15767867 |

Tab.5   Measuring inshore fishing capacity in 1997

| | Inshore fishing caich(t) | Number of engine vessels | | | Labour force | Calculated by DEA | |
|---|---|---|---|---|---|---|---|
| | | Number of vessels | Gross tonnage(t) | Power (kW) | | Capacity utilization | Capacity output(t) |
| Tianjing | 21781 | 995 | 19874 | 35290 | 2,203 | 0.502 | 43366.91 |
| Hebei | 257722 | 9462 | 163094 | 293838 | 32,347 | 0.463 | 556773.3 |
| Liaoning | 1261061 | 31081 | 307576 | 644659 | 124,819 | 0.919 | 1371794 |
| Shanghai | 88084 | 791 | 68524 | 126776 | 3,256 | 1 | 88084 |
| Jiangshu | 699165 | 23023 | 375680 | 743529 | 97,617 | 0.474 | 1473807 |
| Zhejiang | 2777129 | 39664 | 1623524 | 3060147 | 191,177 | 0.864 | 3213051 |
| Fujian | 1878590 | 52443 | 585396 | 1253202 | 216,882 | 0.72 | 2610875 |
| Sandong | 2808271 | 45890 | 629655 | 1179585 | 187,937 | 1 | 2808271 |
| Guangdong | 1847163 | 53073 | 780071 | 2128391 | 213,451 | 0.554 | 3334344 |
| Guangxi | 781429 | 11457 | 205346 | 482261 | 39,180 | 1 | 781429 |
| Hainan | 396386 | 13641 | 186475 | 438518 | 68,495 | 0.477 | 831681.4 |
| Whole country | 12816781 | | | | | | 17113476 |

Tab.6   Measuring inshore fishing capacity in 1998

| | Inshore fishing caich(t) | Number of engine vessels | | | Labour force | Calculated by DEA | |
|---|---|---|---|---|---|---|---|
| | | Number of vessels | Gross tonnage(t) | Power (kW) | | Capacity utilization | Capacity output(t) |
| Tianjing | 23617 | 1016 | 21228 | 37472 | 3142 | 0.345 | 68398.41 |
| Hebei | 299140 | 9082 | 168934 | 294910 | 31792 | 0.485 | 617003.3 |
| Liaoning | 1423564 | 31812 | 304637 | 681033 | 124635 | 0.947 | 1502945 |
| Shanghai | 76245 | 793 | 71878 | 131259 | 3750 | 1 | 76245 |
| Jiangshu | 696698 | 20790 | 357603 | 710101 | 92266 | 0.459 | 1519346 |
| Zhejiang | 3105367 | 39900 | 1751455 | 3371717 | 191348 | 0.926 | 3354163 |
| Fujian | 2011563 | 53665 | 602747 | 1614494 | 191441 | 0.676 | 2973689 |
| Sandong | 3181056 | 45956 | 644779 | 1278860 | 186892 | 1 | 3181056 |
| Guangdong | 1908569 | 53788 | 792230 | 2202765 | 232366 | 0.501 | 3808358 |
| Guangxi | 887329 | 11423 | 202888 | 476255 | 40761 | 1 | 887329 |
| Hainan | 440526 | 13978 | 189741 | 458526 | 68319 | 0.471 | 936098.6 |
| Whole country | 14053674 | | | | | | 18924631 |

Tab.7   Measuring inshore fishing capacity in 1999

| | Inshore fishing caich(t) | Number of engine vessels | | | Labour force | Calculated by DEA | |
|---|---|---|---|---|---|---|---|
| | | Number of vessels | Gross tonnage(t) | Power (kW) | | Capacity utilization | Capacity output(t) |
| Tianjing | 22219 | 1061 | 23033 | 41508 | 3216 | 0.355 | 62564.82 |
| Hebei | 325302 | 9000 | 167166 | 333779 | 31301 | 0.54 | 602790.1 |
| Liaoning | 1376773 | 32468 | 322720 | 762130 | 123764 | 0.809 | 1701542 |
| Shanghai | 79154 | 795 | 66312 | 121821 | 3772 | 1 | 79154 |
| Jiangshu | 668785 | 19811 | 375083 | 733425 | 91256 | 0.456 | 1466619 |
| Zhejiang | 3129758 | 39527 | 1823528 | 3470713 | 186182 | 0.926 | 3378609 |
| Fujian | 2017972 | 52609 | 613018 | 1694731 | 184976 | 0.64 | 3152392 |
| Sandong | 3172260 | 44506 | 601661 | 1263095 | 189041 | 1 | 3172260 |
| Guangdong | 1885442 | 53371 | 799422 | 2190946 | 216781 | 0.489 | 3855609 |
| Guangxi | 888557 | 11753 | 204561 | 493090 | 46087 | 1 | 888557 |
| Hainan | 510910 | 13997 | 193843 | 474926 | 73106 | 0.509 | 1002811 |
| Whole country | 14077132 | | | | | | 19362909 |

Tab.8    Measuring inshore fishing capacity in 2000

| | Inshore fishing caich(t) | Number of engine vessels | | | Labour force | Calculated by DEA | |
|---|---|---|---|---|---|---|---|
| | | Number of vessels | Gross tonnage(t) | Power (kW) | | Capacity utilization | Capacity output(t) |
| Tianjing | 24529 | 1154 | 24591 | 47882 | 3190 | 0.445 | 55151.11 |
| Hebei | 326231 | 8932 | 143679 | 325990 | 27625 | 0.683 | 477601.7 |
| Liaoning | 1334582 | 34367 | 369718 | 831595 | 124663 | 0.759 | 1757495 |
| Shanghai | 60931 | 877 | 66168 | 124137 | 3939 | 0.905 | 67323.94 |
| Jiangshu | 644293 | 19245 | 343053 | 736641 | 86285 | 0.484 | 1331504 |
| Zhejiang | 3192388 | 39385 | 1951625 | 3703671 | 187930 | 1 | 3192388 |
| Fujian | 2029779 | 54832 | 641794 | 1692214 | 190296 | 0.674 | 3011607 |
| Sandong | 2947784 | 45089 | 620115 | 1292370 | 187698 | 1 | 2947784 |
| Guangdong | 1861821 | 57537 | 738049 | 2171835 | 227873 | 0.531 | 3508396 |
| Guangxi | 888417 | 12708 | 216070 | 509611 | 51387 | 1 | 888417 |
| Hainan | 598587 | 14316 | 189507 | 493144 | 74585 | 0.664 | 900842.1 |
| Whole country | 13909342 | | | | | | 18138509 |

Tab.9    Measuring inshore fishing capacity in 2001

| | Inshore fishing caich(t) | Number of engine vessels | | | Labour force | Calculated by DEA | |
|---|---|---|---|---|---|---|---|
| | | Number of vessels | Gross tonnage(t) | Power (kW) | | Capacity utilization | Capacity output(t) |
| Tianjing | 25140 | 1194 | 27095 | 51177 | 2580 | 0.456 | 55177.88 |
| Hebei | 317647 | 8495 | 166691 | 343802 | 27177 | 0.551 | 576392.9 |
| Liaoning | 1347991 | 34518 | 412839 | 859194 | 126358 | 0.829 | 1625809 |
| Shanghai | 60935 | 790 | 65497 | 123366 | 3604 | 0.967 | 63020.47 |
| Jiangshu | 625158 | 16688 | 300366 | 627799 | 90866 | 0.564 | 1107730 |
| Zhejiang | 3109709 | 38707 | 2024096 | 3810467 | 179860 | 1 | 3109709 |
| Fujian | 2010508 | 55035 | 597619 | 1661055 | 196964 | 0.812 | 2476589 |
| Sandong | 2637786 | 43386 | 636517 | 1393995 | 193524 | 1 | 2637786 |
| Guangdong | 1802761 | 55567 | 768278 | 2236958 | 234181 | 0.566 | 3183816 |
| Guangxi | 893069 | 12462 | 231420 | 533599 | 41758 | 1 | 893069 |
| Hainan | 690562 | 14599 | 221299 | 545733 | 78044 | 0.761 | 907423.3 |
| Whole country | 13521266 | | | | | | 16636523 |

Tab.10 Measuring inshore fishing capacity in 2002

| | Inshore fishing caich(t) | Number of engine vessels | | | Labour force | Calculated by DEA | |
|---|---|---|---|---|---|---|---|
| | | Number of vessels | Gross tonnage(t) | Power (kW) | | Capacity utilization | Capacity output(t) |
| Tianjing | 25393 | 1086 | 21571 | 45352 | 2349 | 0.373 | 68089.2 |
| Hebei | 315191 | 8225 | 160540 | 350743 | 27496 | 0.599 | 525777.1 |
| Liaoning | 1302007 | 34499 | 447319 | 890045 | 129437 | 0.822 | 1584575 |
| Shanghai | 49255 | 737 | 47084 | 98514 | 3740 | 0.812 | 60671.92 |
| Jiangshu | 587158 | 14766 | 295678 | 647323 | 78931 | 0.614 | 955716.7 |
| Zhejiang | 3024376 | 36738 | 2066347 | 3869487 | 174739 | 1 | 3024376 |
| Fujian | 1985263 | 53688 | 574613 | 1731531 | 192129 | 0.875 | 2269181 |
| Sandong | 2566052 | 40919 | 649788 | 1441334 | 177225 | 1 | 2566052 |
| Guangdong | 1731780 | 57628 | 815160 | 2315988 | 236617 | 0.538 | 3219116 |
| Guangxi | 862969 | 13920 | 235834 | 592595 | 14674 | 1 | 862969 |
| Hainan | 789035 | 15185 | 242722 | 597519 | 77336 | 0.828 | 953030.1 |
| Whole country | 13238479 | 277391 | 5556656 | 12580431 | 1114673 | 0.769 | 16089554 |

Tab.11 Measuring inshore fishing capacity in 2003

| | Inshore fishing caich(t) | Number of engine vessels | | | Labour force | Calculated by DEA | |
|---|---|---|---|---|---|---|---|
| | | Number of vessels | Gross tonnage(t) | Power (kW) | | Capacity utilization | Capacity output(t) |
| Tianjing | 37105 | 1007 | 30397 | 57780 | 2169 | 0.691 | 53663 |
| Hebei | 101852 | 7509 | 60218 | 198852 | 15570 | 0.463 | 220188 |
| Liaoning | 1394920 | 35376 | 466820 | 1065088 | 131081 | 0.818 | 1704611 |
| Shanghai | 114978 | 762 | 128366 | 126804 | 3526 | 1 | 114978 |
| Jiangshu | 328856 | 14314 | 109507 | 366136 | 70522 | 0.801 | 410523 |
| Zhejiang | 2996821 | 36213 | 2029904 | 3833209 | 168441 | 0.975 | 3029141 |
| Fujian | 2048442 | 52218 | 546421 | 1618168 | 190790 | 1 | 2048442 |
| Sandong | 2593308 | 42277 | 697766 | 1542142 | 165443 | 1 | 2536141 |
| Guangdong | 1819063 | 59745 | 797417 | 2323553 | 244722 | 0.687 | 2950571 |
| Guangxi | 849277 | 14397 | 270714 | 686635 | 38146 | 1 | 851042 |
| Hainan | 707413 | 15138 | 240243 | 618367 | 82735 | 1 | 893845 |
| Whole country | 12992035 | | | | | | 14813145 |

Tab.12 Measuring inshore fishing capacity in 2004

| | Inshore fishing caich(t) | Number of engine vessels | | | Labour force | Calculated by DEA | |
|---|---|---|---|---|---|---|---|
| | | Number of vessels | Gross tonnage(t) | Power (kW) | | Capacity utilization | Capacity output(t) |
| Tianjing | 22874 | 852 | 19038 | 39196 | 2129 | 0.514 | 44523 |
| Hebei | 306410 | 7620 | 139160 | 319617 | 22428 | 0.854 | 358864 |
| Liaoning | 1265002 | 35306 | 465842 | 967848 | 118124 | 0.799 | 1583232 |
| Shanghai | 37616 | 637 | 41779 | 88621 | 2239 | 0.861 | 43694 |
| Jiangshu | 571197 | 14554 | 282602 | 650368 | 81238 | 0.593 | 962477 |
| Zhejiang | 2935908 | 34649 | 1960640 | 3747543 | 158138 | 1.000 | 2935908 |
| Fujian | 2064841 | 54955 | 642532 | 1895572 | 194004 | 0.792 | 2607530 |
| Sandong | 2479510 | 39532 | 610986 | 1318917 | 162372 | 1.000 | 2479510 |
| Guangdong | 1586573 | 59573 | 774913 | 2265699 | 235263 | 0.505 | 3144760 |
| Guangxi | 799488 | 14558 | 273762 | 698677 | 38230 | 1.000 | 799488 |
| Hainan | 990365 | 15705 | 259916 | 686264 | 77697 | 0.975 | 1016056 |
| Whole country | 13059784 | | | | | | 14633268 |

Tab.13 Measuring inshore fishing capacity in 2005

| | Inshore fishing caich(t) | Number of engine vessels | | | Labour force | Calculated by DEA | |
|---|---|---|---|---|---|---|---|
| | | Number of vessels | Gross tonnage(t) | Power (kW) | | Capacity utilization | Capacity output(t) |
| Tianjing | 26573 | 729 | 17106 | 34664 | 1682 | 0.718 | 37023 |
| Hebei | 306962 | 7237 | 137905 | 312232 | 25363 | 0.688 | 445884 |
| Liaoning | 1264873 | 34254 | 507001 | 1005568 | 114130 | 0.732 | 1728794 |
| Shanghai | 34469 | 680 | 43649 | 91342 | 2499 | 0.704 | 48934 |
| Jiangshu | 562458 | 13873 | 258625 | 632295 | 66766 | 0.632 | 889872 |
| Zhejiang | 2911502 | 33943 | 1929322 | 3675540 | 160304 | 1 | 2911502 |
| Fujian | 2047912 | 55042 | 632084 | 1912688 | 186693 | 0.835 | 2454010 |
| Sandong | 2422822 | 38556 | 629916 | 1409255 | 157190 | 1 | 2422822 |
| Guangdong | 1594244 | 57704 | 769604 | 2219148 | 225041 | 0.534 | 2986128 |
| Guangxi | 843286 | 14385 | 272649 | 697422 | 38312 | 1 | 843286 |
| Hainan | 1079799 | 19382 | 275397 | 777243 | 94771 | 1 | 1079799 |
| Whole country | 13094900 | | | | | | 15848054 |